\documentclass[twoside]{article}

\usepackage{graphicx}

\usepackage{cmpj2e}

\newcommand{\op}[1]{%
    \fontdimen12\textfont3=2pt\fontdimen12\scriptfont3=1.4pt%
    \!\null\mathop{\vphantom{#1}\smash{#1}}\limits_{\sim}\null\!}
\newcommand{\xref}[1]{\protect\ref{#1}}
\newcommand{\figref}[1]{Fig.~\protect\ref{#1}}
\newcommand{\fmref}[1]{(\protect\ref{#1})}

\newcommand{\vek}[1]{{\!\vec{\,#1}}}

\title[]%
{Magnetic response of magnetic molecules with
  non-collinear local $d$-tensors%
\thanks{jschnack@physik.uni-bielefeld.de}}
\author[]{J{\"u}rgen Schnack\refaddr{label1}}
\addresses{
\addr{label1} Universit{\"a}t Bielefeld, Fakult{\"a}t f{\"u}r Physik, Postfach 100131, D-33501 Bielefeld, Germany
}
\begin{document}

\maketitle

\begin{abstract}
Investigations of molecular magnets are driven both by
prospective applications in future storage technology or quantum
computing as well as by fundamental questions. Nowadays
numerical simulation techniques and computer capabilities make
it possible to investigate spin Hamiltonians with realistic
arrangements of local anisotropy tensors. In this contribution I
will discuss the magnetic response of a small spin system with
special emphasis on non-collinear alignments of the local
anisotropy axes.
\keywords Magnetic Molecules; Single-Ion Anisotropy; Exact Diagonalization
\pacs 75.50.Xx; 75.10.Jm; 75.40.Cx; 75.40.Mg
\end{abstract}

%%%%%%%%%%%%%%%%%%%%%%%%%%%%%%%%%%%%%%%%%%%%%%%%%%%%%%%%%%%%%%%%%%%%%%%%
\section{Introduction}

Since the early investigations of Mn$_{12}$-acetate
\cite{Lis:ACB80} single molecule magnets (SMM) are at the heart
of the investigations of magnetic molecules worldwide. This is
due to their properties which are governed by the anisotropy
barrier, as there are slow relaxation of the magnetization as
well as spin tunneling through the barrier
\cite{ThB:JLTP98,CGJ:PRL00,PhysRevB.61.1286,PhysRevB.62.15026}.
From a chemical point of view it is striking that until recently
no compound could be synthesized with a higher anisotropy
barrier which moved hopes of easy application into the more
distant future. Theoretical estimates pointed out that this
might be due to very general reasons \cite{Wal:IC07,RCC:CC08}.  A
recently synthesized manganese compound put an end to more than
20 years of search \cite{MVW:JACS07}.

On the theory side numerically exact evaluations of spin
Hamiltonians including anisotropic terms turned out to be
limited to rather small systems such as for instance an
antiferromagnetically coupled Ni$_4$ compound
\cite{PBS:PT05,SBL:PRB06,KMS:CCR09,NHB:09}, which was
investigated in great detail or another but ferromagnetically
coupled Ni$_4$ \cite{KST:PRB08}. With great numerical effort the
exchange constants of Mn$_{12}$-acetate could be determined
\cite{CSO:PRB04}, but usually one resorts to models where the
interacting spin system is replaced by one large spin in its
effective ligand field (giant spin approximation).  But thanks
to the technical progress detailed numerical studies of larger
molecules are affordable nowadays. First examples are given by
the simulation of the new Mn$_6$ compound with record anisotropy
barrier \cite{CGS:PRL08} and by the investigation of a Mn$_6$Fe
compound \cite{GHK:IC09}. Whereas Ref.~\cite{CGS:PRL08} restricts
its parameter space to collinear local anisotropy axes,
Ref.~\cite{GHK:IC09} considers non-collinear ones.

In this contribution the recently developed procedures will be
used to discuss the influence of non-collinear anisotropy axes
on the magnetic response of a fictitious molecular compound. In
order not to make the situation too complicated I will restrict
the discussion to a triangular arrangement of equal spins of
$s=1$ with a molecular $C_3$ symmetry. Then the magnetization
will be discussed for three typical scenarios given by the
possible ratios of exchange and anisotropy: $|J| > |d|$, $|J| =
|d|$, and $|J| < |d|$.

%%%%%%%%%%%%%%%%%%%%%%%%%%%%%%%%%%%%%%%%%%%%%%%%%%%%%%%%%%%%%%%%%%%%%%%%
\section{Hamiltonian and evaluation}
\subsection{Hamiltonian}

It turns out that many magnetic molecules can be well described
by a Heisenberg model which takes the super exchange between the
moments of the (typically $3d$) ions into account. In this
\emph{strong coupling limit} anisotropic terms appear as
perturbations and are thus often taken care of by perturbation
theory. But for single molecule magnets, that might e.g.
contain Mn(III) or V(III) ions, the situation might be reversed
or anisotropy and exchange are at least of the same
order. Therefore, I would like to treat both on the same footing
right from the beginning. In the following Hamiltonian, 
%--------------------------------------------------------
\begin{eqnarray}
\label{E-2-1}
\op{H}(\vek{B})
&=&
-
\sum_{i,j}\;
{J}_{ij}
\op{\vek{s}}_i \cdot \op{\vek{s}}_j
+
\sum_{i}\;
d_i\, \big(\vek{e}_i\cdot\op{\vek{s}}_i\big)^2
+
\mu_B\, \vek{B}\cdot\,
\sum_{i}\;
{\mathbf g}_i\cdot
\op{\vek{s}}_i
\ ,
\end{eqnarray}
%--------------------------------------------------------
the first term models the isotropic Heisenberg exchange
interaction. A negative ${J}_{ij}$ corresponds to an
antiferromagnetic coupling of spins at sites $i$ and $j$. The
second term models the local anisotropy tensors by their major
principal axis which points along $\vek{e}_i$. Depending on the
sign of $d_i$ this is an easy ($d_i<0$) or hard ($d_i>0$)
axis. The neglected two other principal axes of the local
${\mathbf d}$--tensor ($e$-terms) are usually much smaller.
The last term provides the interaction with the applied magnetic
field. Here ${\mathbf g}_i$ is the local $g$-tensor, which might
also be anisotropic.

%===================    figure   =================================
\begin{figure}[ht!]
\centering
\includegraphics[clip,width=65mm]{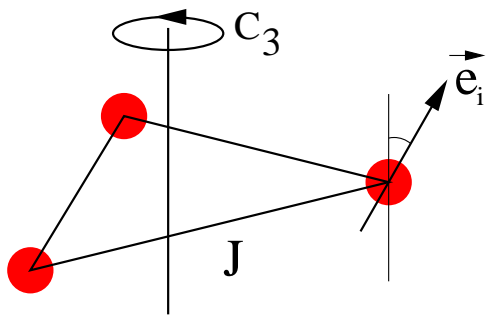}
\caption{Configuration of the fictitious $C_3$-symmetric
  molecule discussed in this article. The remaining parameters
  of the model are the exchange coupling $J$, the strength of
  the local anisotropy $d$, and the angle $\vartheta$ of the local
  anisotropy axes with respect to the $C_3$ axis.}
\label{F-1}
\end{figure}
%===================    figure   =================================

Since in this contribution the influence of the non-collinearity
of the local anisotropy axes shall be discussed, the Hamiltonian
will be restricted to a simplified spin system, that is depicted
in \figref{F-1}. Three spins $s=1$ of a fictitious
$C_3$-symmetric molecule will be considered that interact with
one and the same exchange $J$. The local anisotropy axes are
also related by the $C_3$ symmetry, so that the anisotropy can
be modeled by one strength $d$ and one azimuthal angle
$\vartheta$. The $g$-tensors are assumed to be isotropic and of
value 2.

%%%%%%%%%%%%%%%%%%%%%%%%%%%%%%%%%%%%%%%%%%%%%%%%%%%%%%%%%%%%%%%%%%%%%%%%
\subsection{Evaluation}

Since the various parts of Hamiltonian \fmref{E-2-1} do not
commute in general all eigenvalues and eigenvectors have to be
calculated for each magnetic field $\vec{B}$, i.e. for each
strength and direction of the field. The magnetization is then
evaluated using all eigenvectors. I would like to mention that
it is also possible to obtain the magnetization without using
eigenvectors by numerical differentiation of the energy
eigenvalues $E_\nu(\vec{B})$, which are functions of $\vec{B}$.

Very often the investigated substance is only available as a
powder. Then an orientational average has to be performed in
order to be able to compare to experiments.  Since this average
cannot be performed analytically one sums over a finite set of
special directions. In our evaluations we use for each absolute
value $B=|\vec{B}|$ special sets of directions given by points
on the unit sphere. These sets are called Lebedev-Laikov grids
\cite{LeL:DAN99}. The directions, which contribute with various
weights, and their total number are chosen such that the angular
integration of polynomials $x^k \times y^l \times z^m$, where
$k+l+m \le 131$, can be performed with a relative accuracy of
$2^{-14}$ \cite{LeL:DAN99}. In our evaluations we normally use a
Lebedev-Laikov grid with 50 points.  These points can be
generated with publicly available software
\cite{LLGrid}. Equivalently one may cover the unit sphere with
the vertices of regular bodies such as an octahedron, a cube, or
a dodecahedron. For practically all examples we evaluated in the
past we can say that averaging over the directions given by the
vertices of a dodecahedron is as accurate as averaging over a
Lebedev-Laikov grid with 50 points. On the contrary, averaging
only over a smaller number of directions like $\pm x$, $\pm y$,
and $\pm z$ is insufficient.

%%%%%%%%%%%%%%%%%%%%%%%%%%%%%%%%%%%%%%%%%%%%%%%%%%%%%%%%%%%%%%%%%%%%%%%%
\section{Anisotropy versus coupling}

In the following main part the magnetic response of the spin
system shown in \figref{F-1} will be discussed for three
scenarios: $|J| > |d|$, $|J| = |d|$, and $|J| < |d|$.  In most
cases the figures show powder averaged magnetizations for
$\vartheta=0, 10, 20, 30, 40, 50, 60, 70, 80, 90$ degrees.
Solid curves correspond to $\vartheta=0, 10, 20, 30$ degrees,
dashed ones to $\vartheta=40, 50, 60, 70$ degrees, and
dashed-dotted curves depict $\vartheta=80, 90$ degrees.  In
order to pronounce the details all magnetization curves are
shown at a low temperature of $T=0.1 |J|$.

%%%%%%%%%%%%%%%%%%%%%%%%%%%%%%%%%%%%%%%%%%%%%%%%%%%%%%%%%%%%%%%%%%%%%%%%
\subsection{Strong coupling limit}

%===================    figure   =================================
\begin{figure}[ht!]
\centering
\includegraphics[clip,width=65mm]{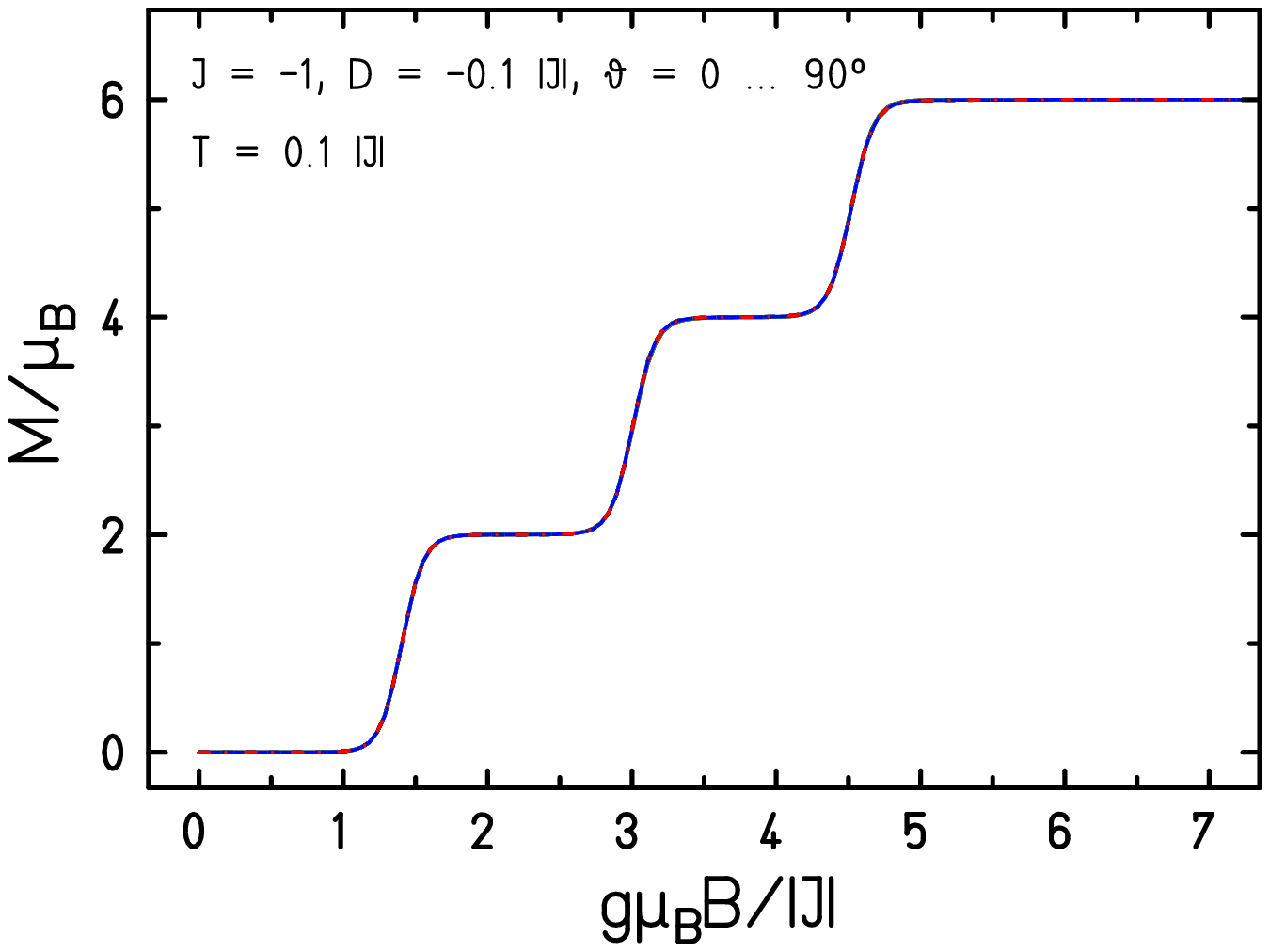}
\quad
\includegraphics[clip,width=65mm]{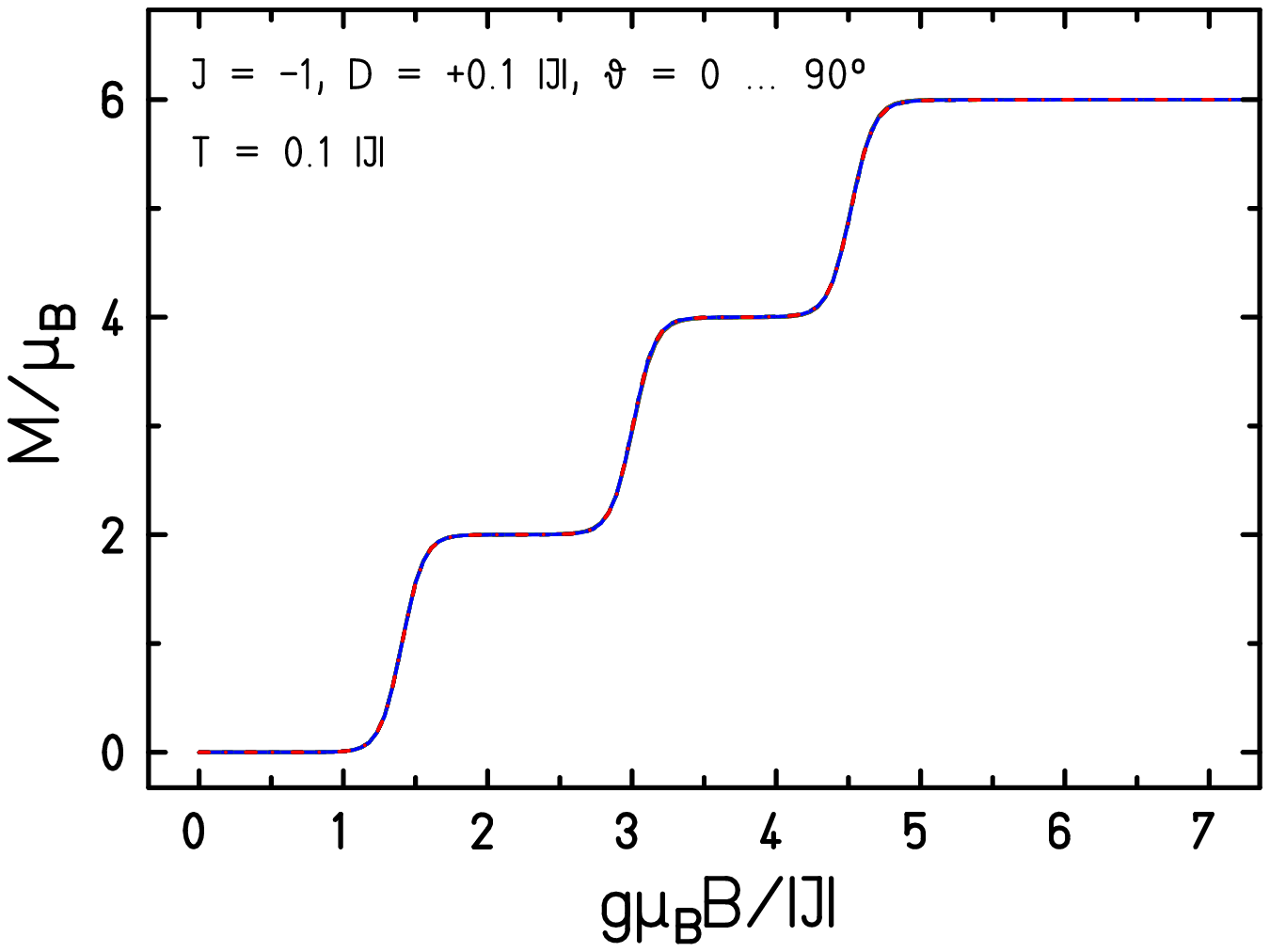}
\caption{Magnetization for antiferromagnetic coupling $J=-1$ and
  easy axis anisotropy $d=-0.1 |J|$ (l.h.s.) as well as hard axis
  anisotropy $d=+0.1 |J|$ (r.h.s.). $\vartheta=0, 10, 20, 30, 40, 50, 60, 70, 80, 90$
  degrees.} 
\label{F-2}
\end{figure}
%===================    figure   =================================

%===================    figure   =================================
\begin{figure}[ht!]
\centering
\includegraphics[clip,width=65mm]{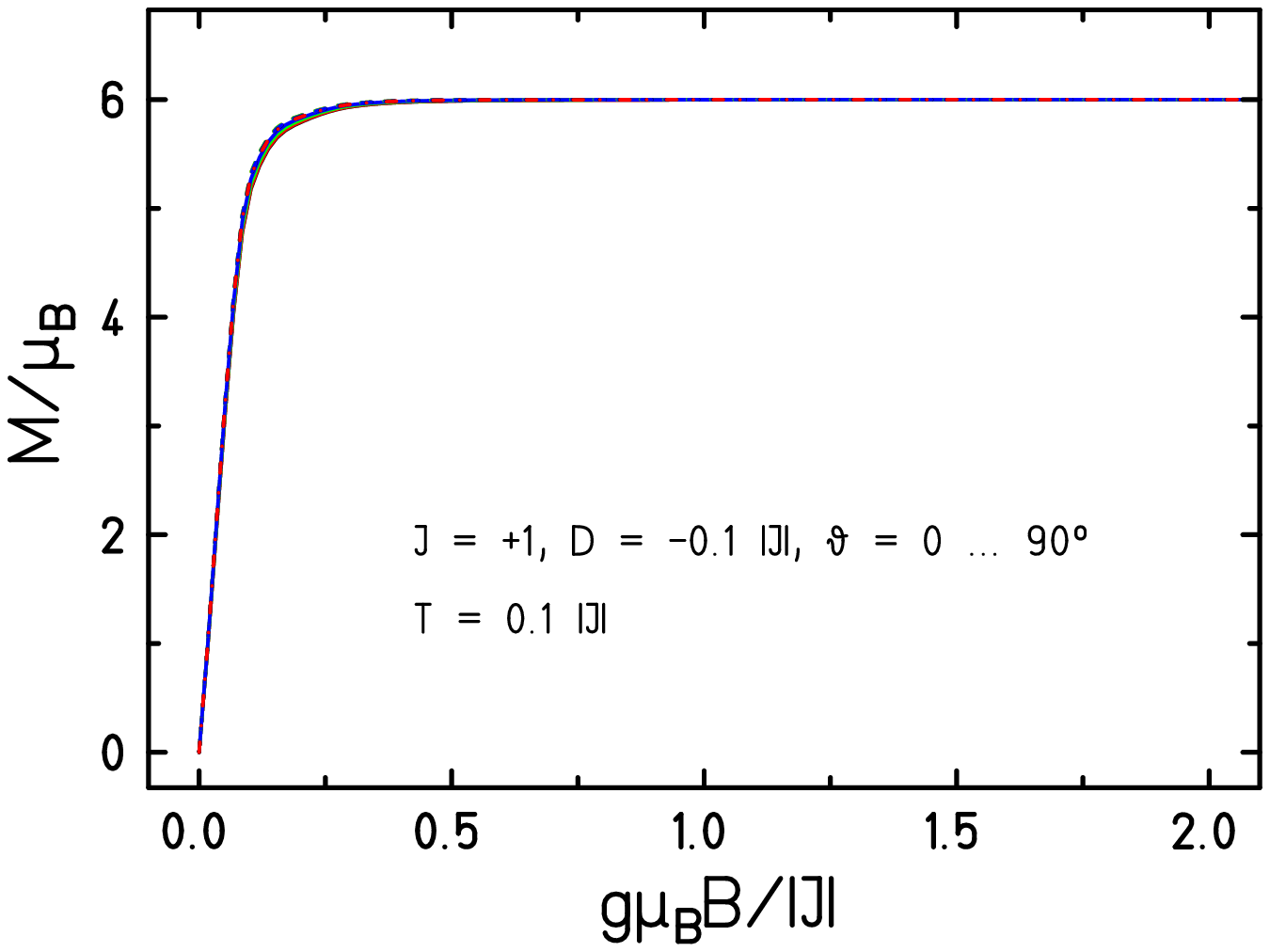}
\quad
\includegraphics[clip,width=65mm]{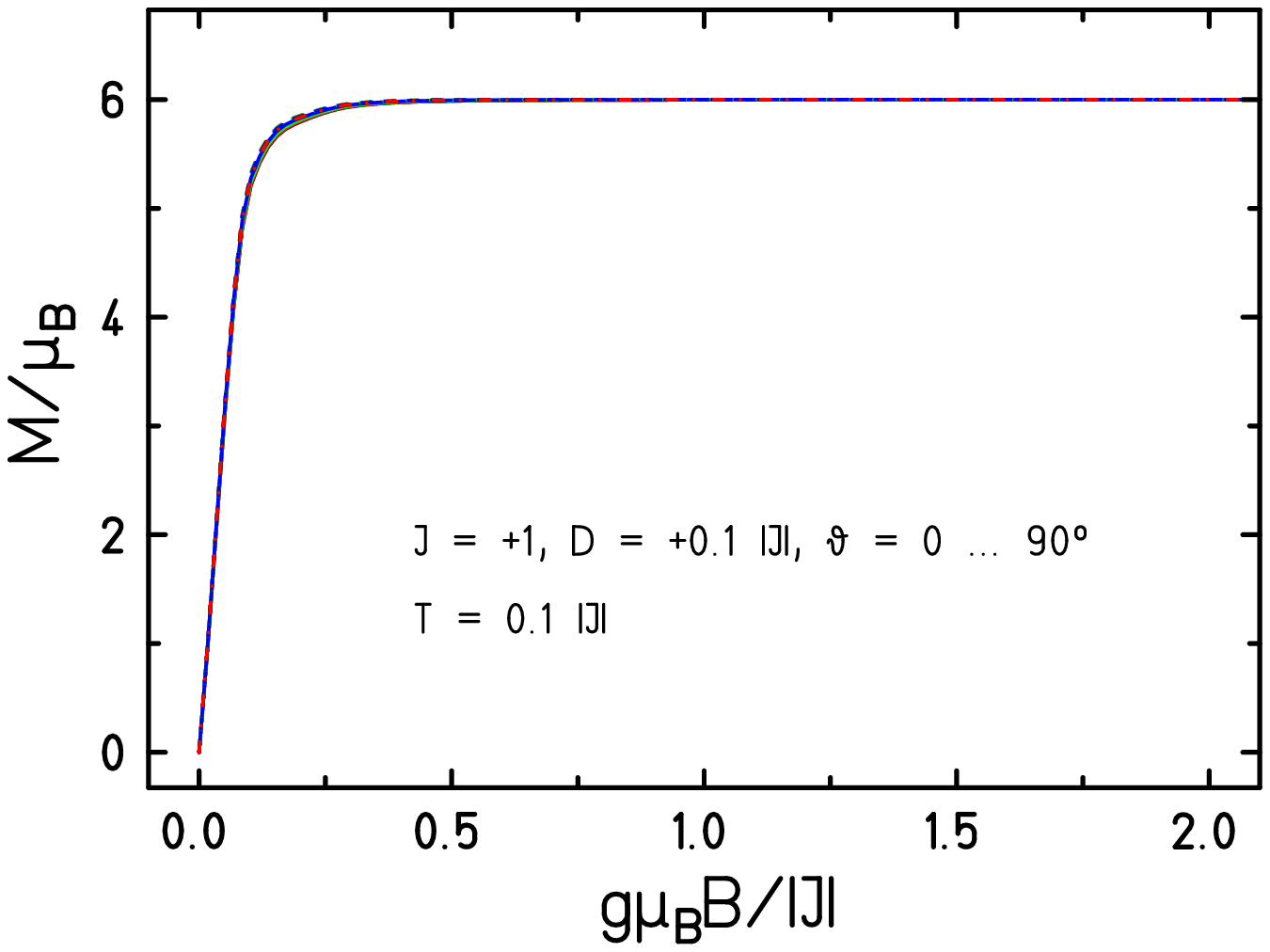}
\caption{Magnetization for ferromagnetic coupling $J=+1$ and
  easy axis anisotropy $d=-0.1 |J|$ (l.h.s.) as well as hard
  axis anisotropy $d=+0.1 |J|$ (r.h.s.). $\vartheta=0, 10, 20, 30,
  40, 50, 60, 70, 80, 90$ degrees.}
\label{F-3}
\end{figure}
%===================    figure   =================================

In the strong coupling limit the exchange energy is much bigger
than the anisotropy energy, i.e. $|J| > |d|$. This is the case
for many molecules, e.g. antiferromagnetically coupled iron or
chromium rings
\cite{TDP:JACS94,PDK:JMMM97,JJL:PRL99,VSG:CEJ02,Waldmann:EPL02,CVG:PRB03,AGC:PRB03,EnL:PRB06,SWH:PRL07,FKK:PRB09}.
These systems thus have in common that a description in terms of
a plain Heisenberg Hamiltonian often provides a very good
approximation
\cite{BSS:JMMM00,BSS:JMMM00:B,Schnack:PRB00,BHS:PRB03}. If
needed the anisotropic terms can then be included via
perturbation theory.

In Fig.~\xref{F-2} and \xref{F-3} the magnetization curves are
shown for combinations of anti-/ferromagnetic coupling $J=\mp1$
and easy/hard axis anisotropy $d=\mp0.1 |J|$. In the
antiferromagnetic case (\figref{F-2}) the low-temperature
magnetization curve is a staircase due to the successive
ground-state level crossings in the growing magnetic field
whereas in the ferromagnetic case the magnetization curve is
practically given by the Brillouin function of the ground state
spin $S=3$. Since the anisotropy is small in magnitude its
angular variation is not noticeable.

%%%%%%%%%%%%%%%%%%%%%%%%%%%%%%%%%%%%%%%%%%%%%%%%%%%%%%%%%%%%%%%%%%%%%%%%
\subsection{Intermediate coupling}

%===================    figure   =================================
\begin{figure}[ht!]
\centering
\includegraphics[clip,width=65mm]{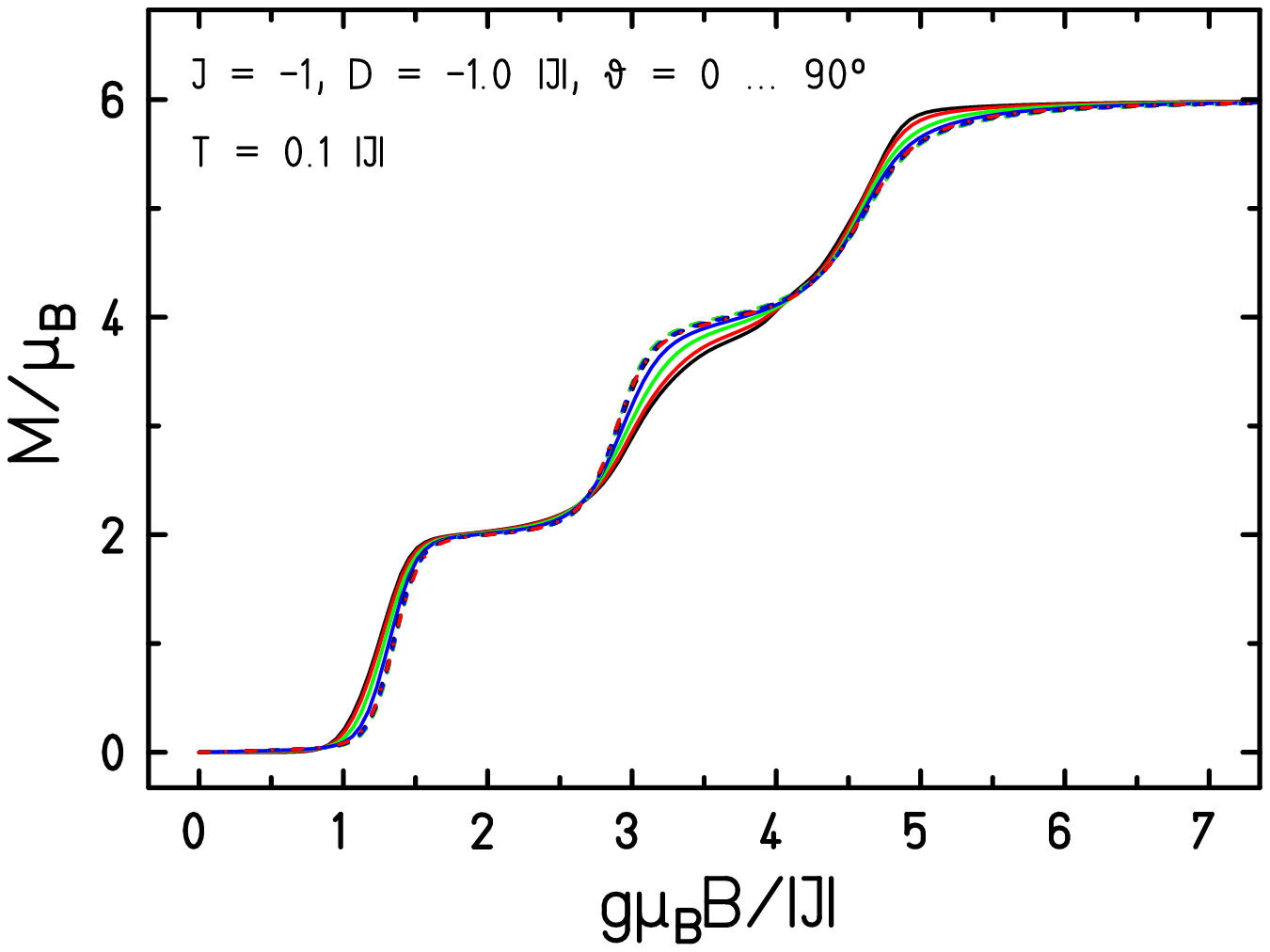}
\quad
\includegraphics[clip,width=65mm]{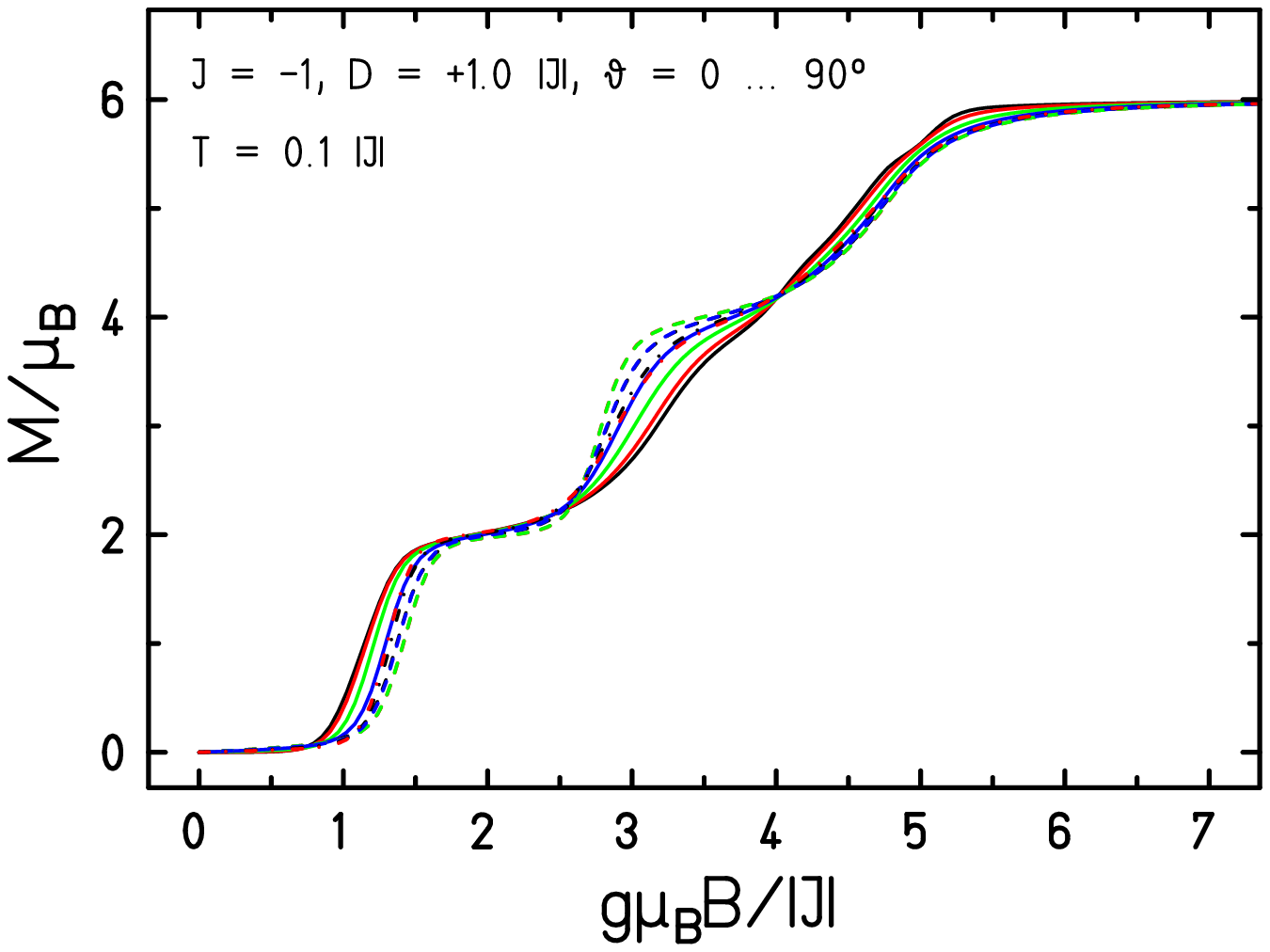}
\caption{Magnetization for antiferromagnetic coupling $J=-1$ and
  easy axis anisotropy $d=-1 |J|$ (l.h.s.) as well as hard axis
  anisotropy $d=+1 |J|$ (r.h.s.). $\vartheta=0, 10, 20, 30, 40, 50, 60, 70, 80, 90$
  degrees.} 
\label{F-4}
\end{figure}
%===================    figure   =================================

%===================    figure   =================================
\begin{figure}[ht!]
\centering
\includegraphics[clip,width=65mm]{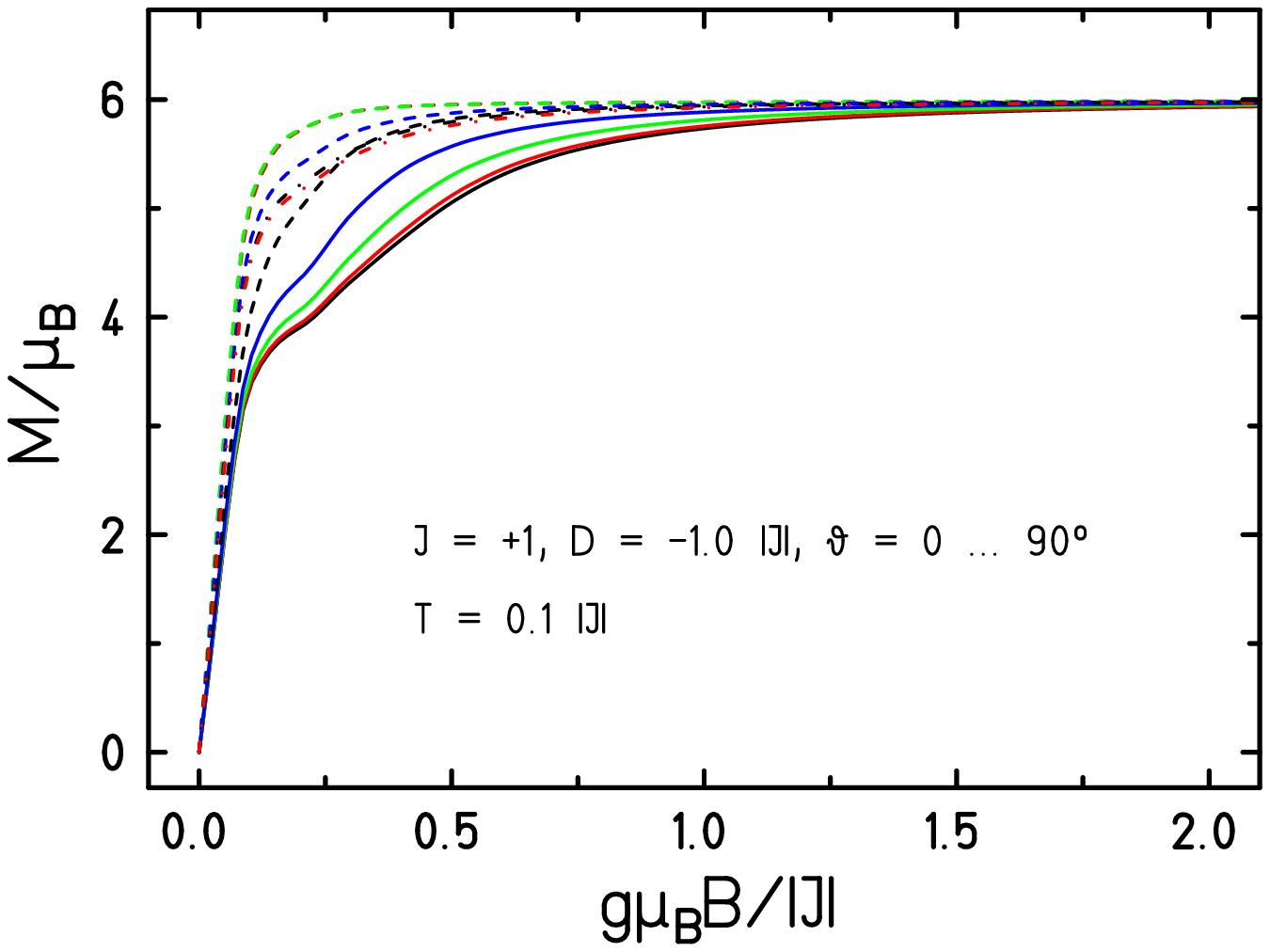}
\quad
\includegraphics[clip,width=65mm]{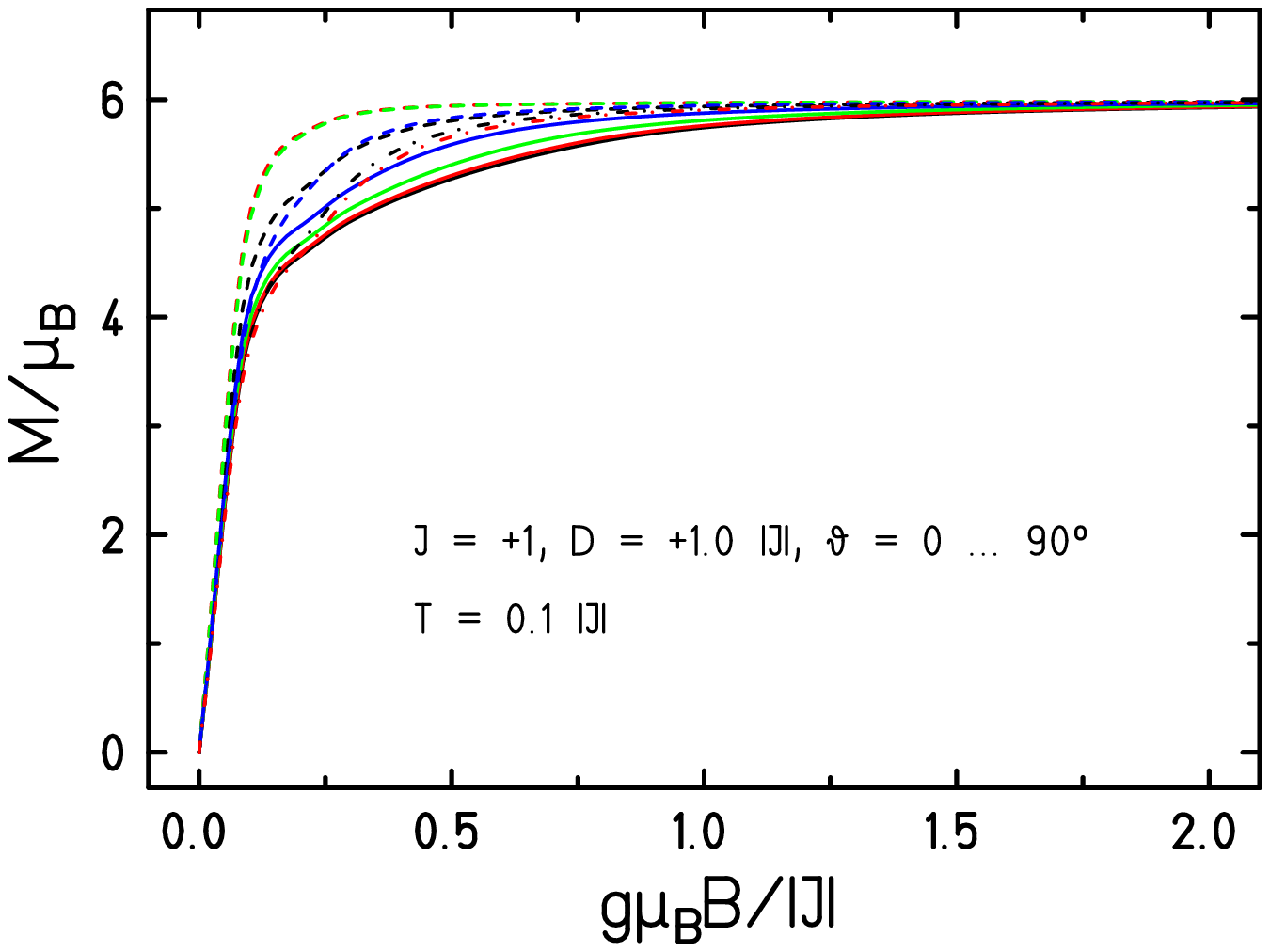}
\caption{Magnetization for ferromagnetic coupling $J=+1$ and
  easy axis anisotropy $d=-1 |J|$ (l.h.s.) as well as hard axis
  anisotropy $d=+1 |J|$ (r.h.s.). $\vartheta=0, 10, 20, 30, 40, 50, 60, 70, 80, 90$
  degrees.} 
\label{F-5}
\end{figure}
%===================    figure   =================================

In the intermediate coupling regime exchange and anisotropy are
of the same order. In the examples presented in Figs.~\xref{F-4}
and \xref{F-5} they have the same amplitude $|J| = |d|$. Looking
at the figures one easily notices that the now larger anisotropy
smears out the angle averaged magnetization
curves. Nevertheless, the underlying structure -- steps for the
antiferromagnetic case and Brillouin function like for the
ferromagnetic case -- is still clearly visible. Interestingly
the structure of the magnetization curve is most similar to the
strong coupling limit for intermediate angles $\vartheta$ around
$50^\circ \dots 60^\circ$.

In detail the curves follow the following trend with varying
$\vartheta$. At small $\vartheta$ (solid curves), i.e. close to
the collinear configuration, the influence of the anisotropy is
strong since it acts cooperatively. Then for intermediate
$\vartheta$ (dashed curves) the influence of the anisotropy is
weakest compared to the strong coupling limit. This is
understandable if one recalls that
$\vartheta=\mbox{arccos}(1/\sqrt{3})=54,73^\circ$ is the angle
of a perfect octahedral alignment of the anisotropy axes,
i.e. the three axis have a pairwise angle of $90^\circ$. In this
configuration the effect of the anisotropic terms on multiplets
of the Heisenberg Hamiltonian is either canceled or rather
small. This explains why both at $50^\circ$ and $60^\circ$
anisotropy effects are hardly visible.
For $\vartheta > 60^\circ$ (dashed-dotted curves) the influence
of the anisotropy rises again.

%%%%%%%%%%%%%%%%%%%%%%%%%%%%%%%%%%%%%%%%%%%%%%%%%%%%%%%%%%%%%%%%%%%%%%%%
\subsection{Weak coupling limit}

%===================    figure   =================================
\begin{figure}[ht!]
\centering
\includegraphics[clip,width=65mm]{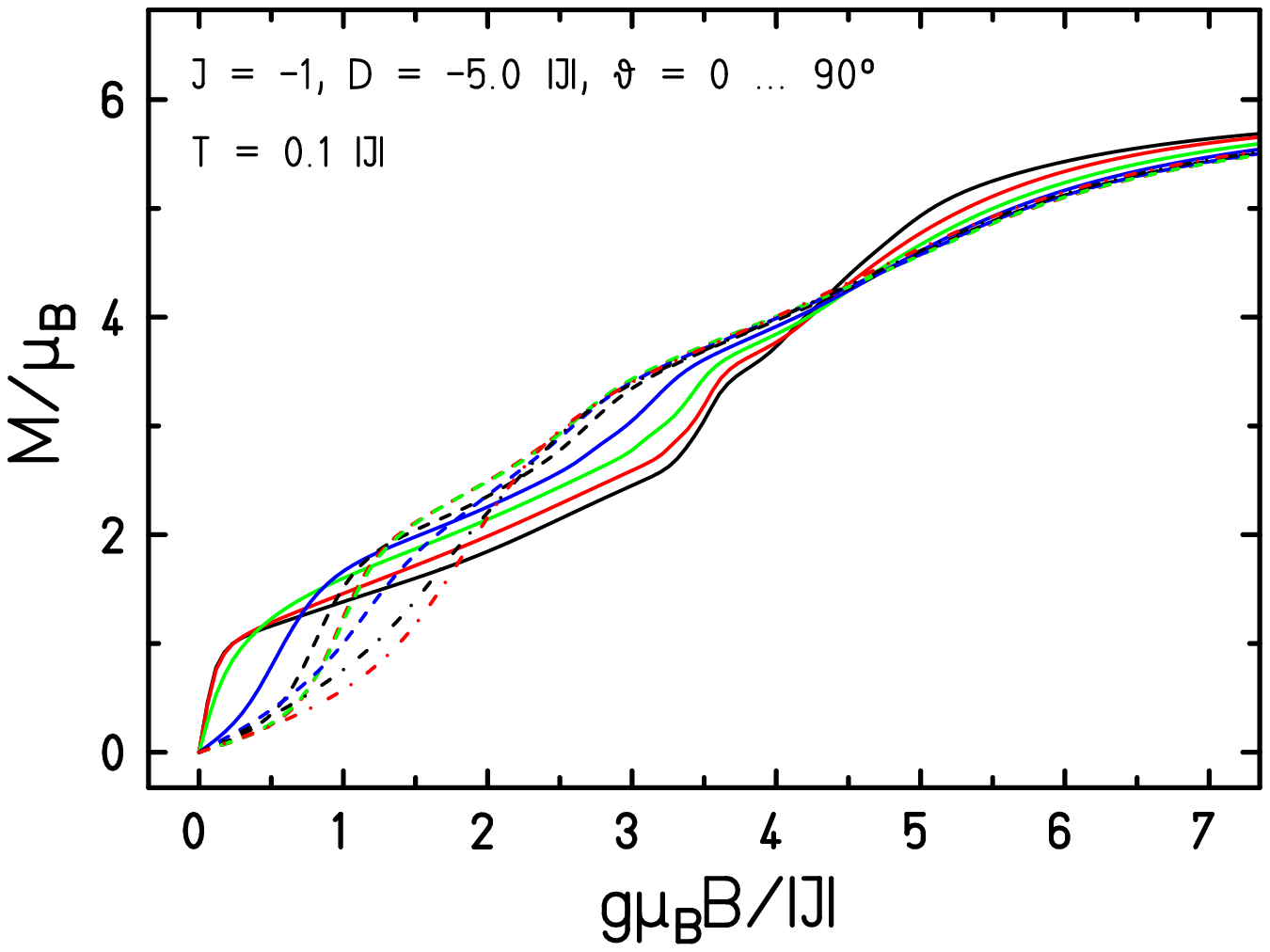}
\quad
\includegraphics[clip,width=65mm]{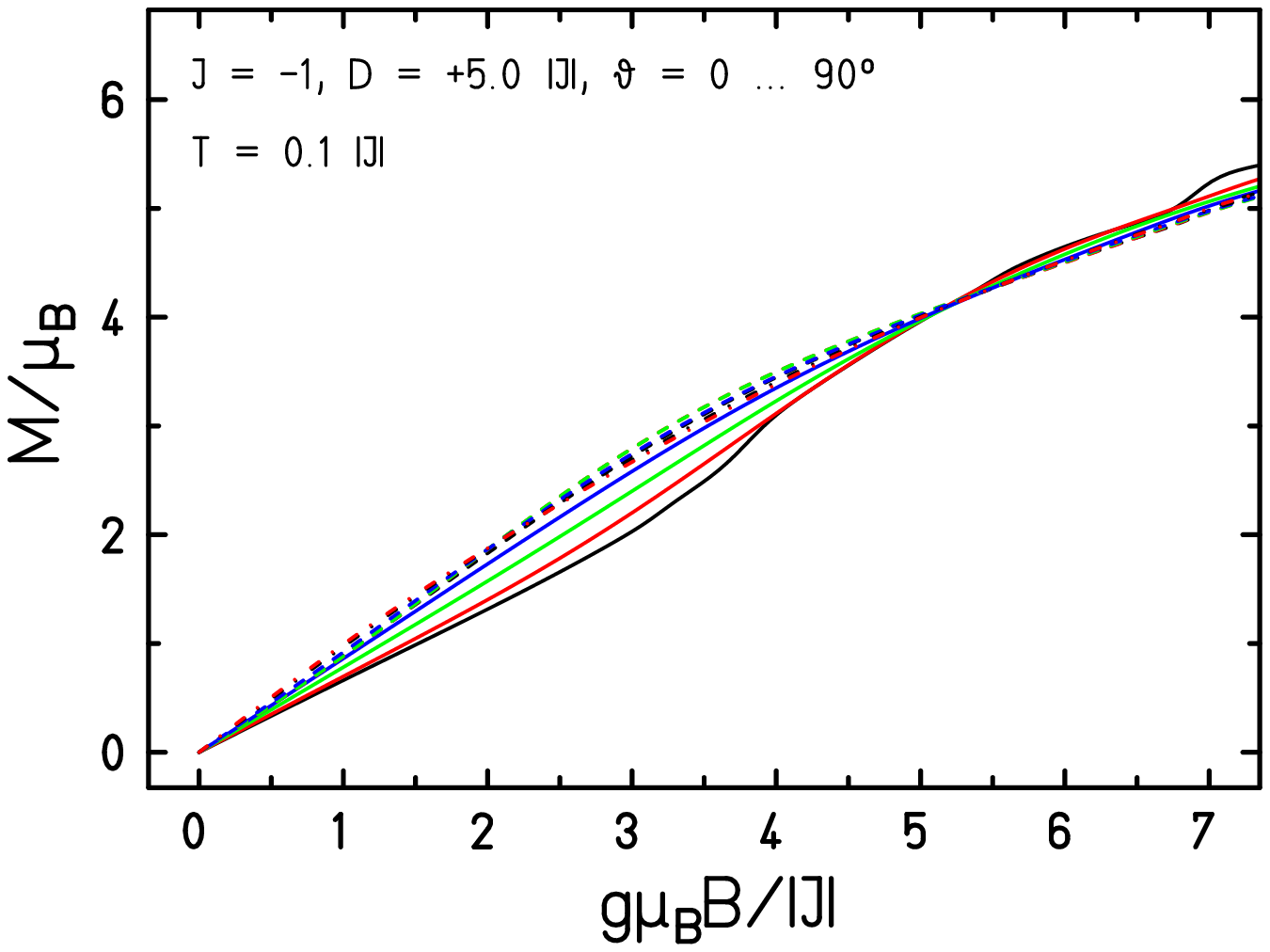}
\caption{Magnetization for antiferromagnetic coupling $J=-1$ and
  easy axis anisotropy $d=-5 |J|$ (l.h.s.) as well as
  hard axis anisotropy $d=+5 |J|$ (r.h.s.). $\vartheta=0, 10,
  20, 30, 40, 50, 60, 70, 80, 90$ degrees.}
\label{F-6}
\end{figure}
%===================    figure   =================================

%===================    figure   =================================
\begin{figure}[ht!]
\centering
\includegraphics[clip,width=65mm]{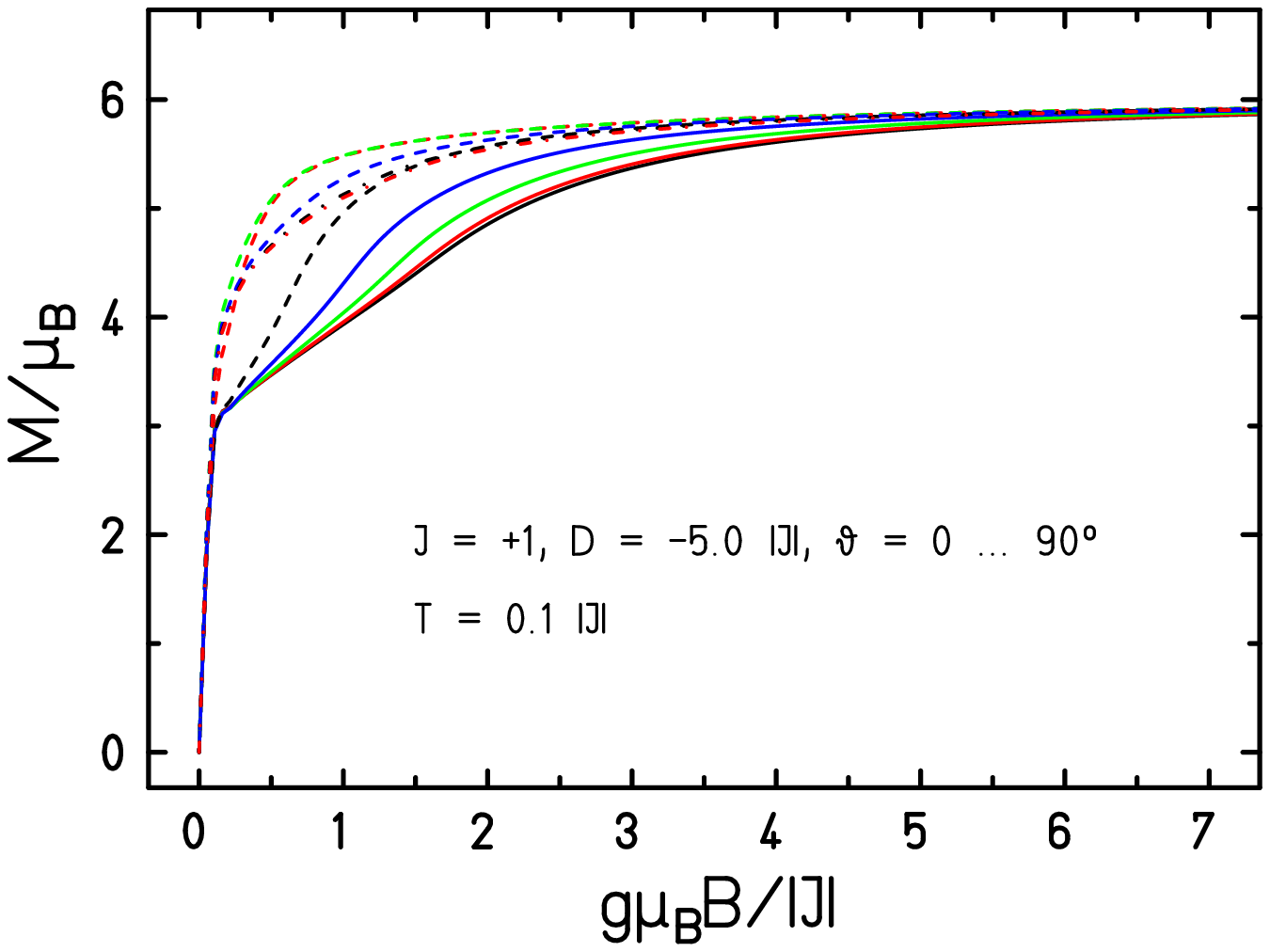}
\quad
\includegraphics[clip,width=65mm]{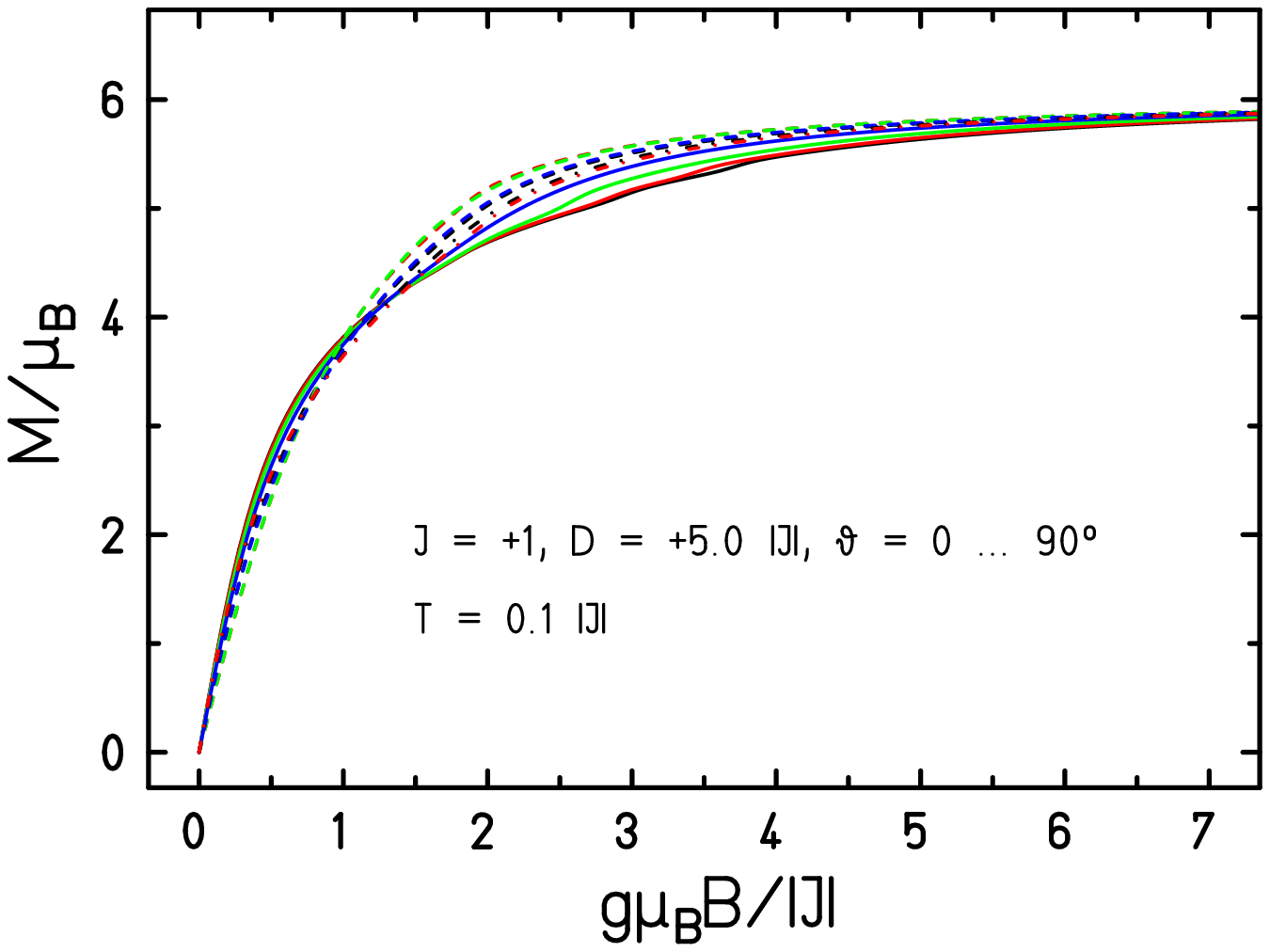}
\caption{Magnetization for ferromagnetic coupling $J=+1$ and
  easy axis anisotropy $d=-5 |J|$ (l.h.s.) as well as
  hard axis anisotropy $d=+5 |J|$ (r.h.s.). $\vartheta=0, 10,
  20, 30, 40, 50, 60, 70, 80, 90$ 
  degrees.}
\label{F-7}
\end{figure}
%===================    figure   =================================

The weak coupling limit, for which the exchange energy is (much)
smaller than the anisotropy energy, i.e. $|J| < |d|$, is
interesting since this case seems to be relevant for an
improvement of single molecule magnets (SMM). The curves in
Figs.~\xref{F-6} and \xref{F-7} show that the effect of the
anisotropy is -- as expected -- strong, and in the case of
antiferromagnetic coupling (\figref{F-6}) very strong. Here one
can say, that a hard axis anisotropy together with
antiferromagnetic coupling (r.h.s. of \figref{F-6}) completely
destroys any structure of the magnetization curve. A (meta-)
stable magnetic ground state that is separated by an anisotropy
barrier from its counterpart of opposite magnetization is either
not created or the low-lying spectrum is so dense that the
mechanism of a barrier breaks down. Zero-field split multiplets
do not exist any more and are thus an inadequate picture for
such a situation.  In the antiferromagnetic case with easy axis
(l.h.s. of \figref{F-6}) one can at least for almost collinear
alignments of the anisotropy axes obtain a non-vanishing ground
state moment that is stabilized by anisotropy.

%===================    figure   =================================
\begin{figure}[ht!]
\centering
\includegraphics[clip,width=65mm]{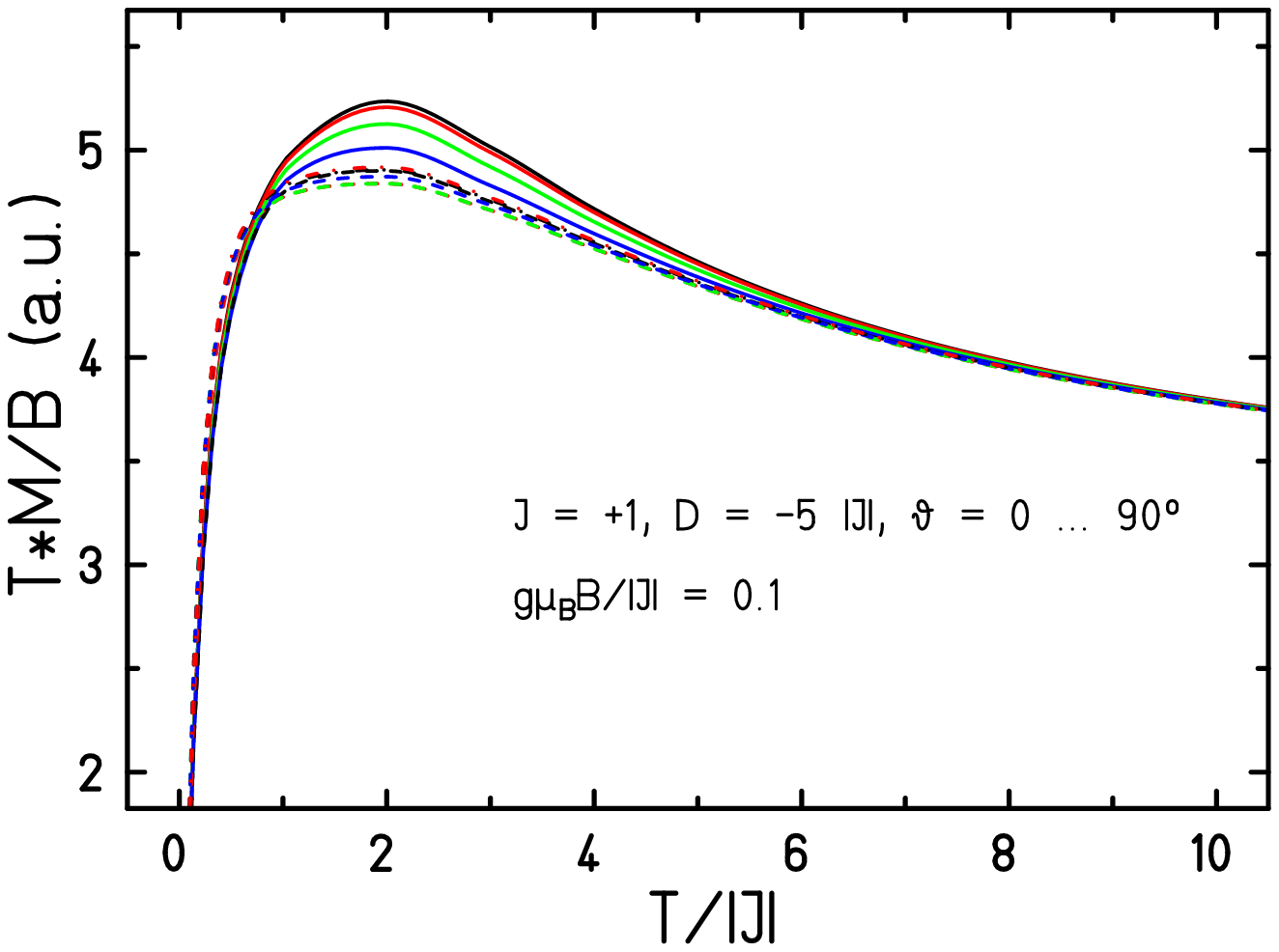}
\quad
\includegraphics[clip,width=65mm]{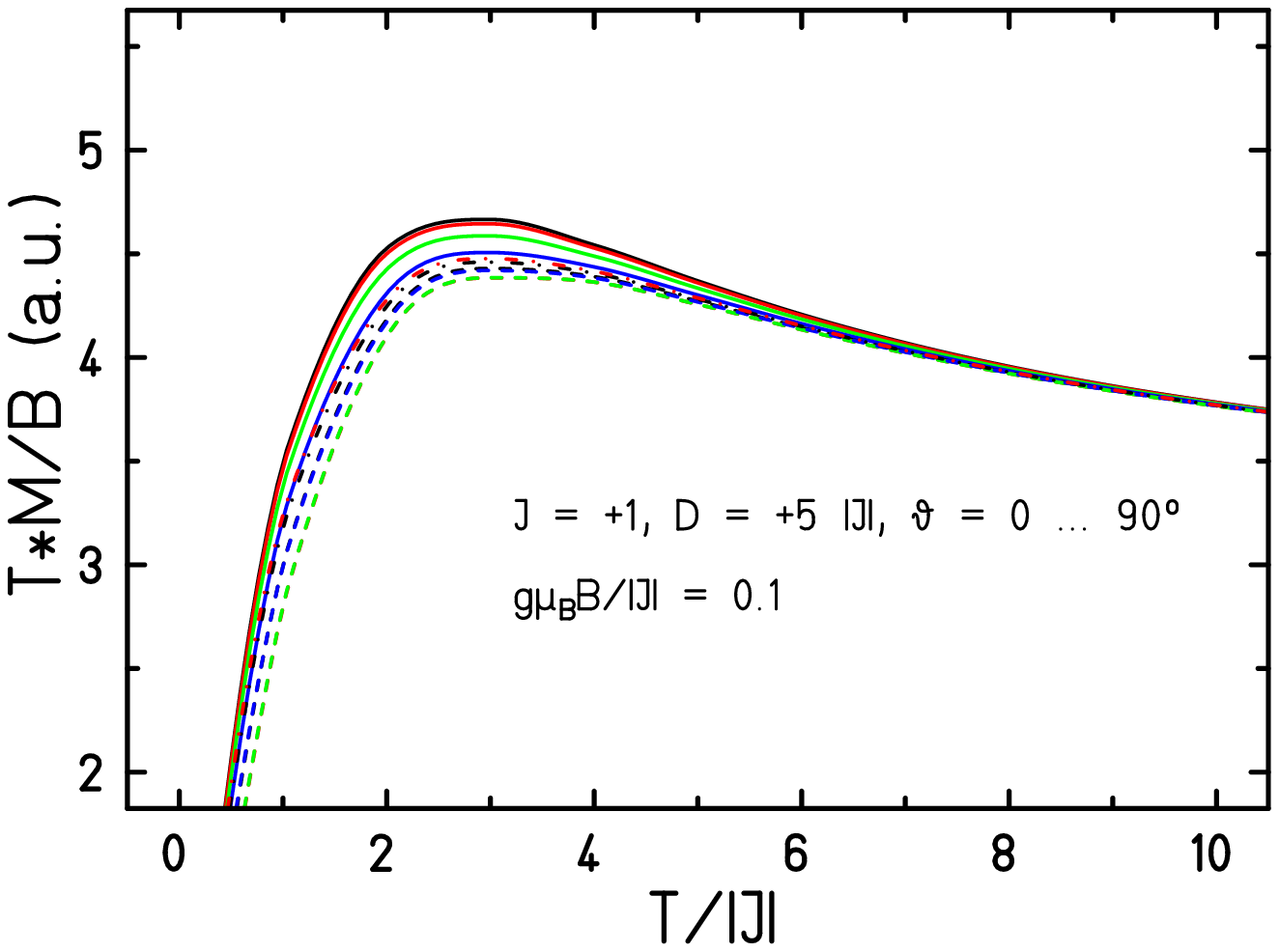}
\caption{Magnetization as a function of temperature for
  $g\mu B/|J|=0.1$ for ferromagnetic coupling $J=+1$ and 
  easy axis anisotropy $d=-5 |J|$ (l.h.s.) as well as
  hard axis anisotropy $d=+5 |J|$ (r.h.s.). $\vartheta=0, 10,
  20, 30, 40, 50, 60, 70, 80, 90$ 
  degrees.}
\label{F-8}
\end{figure}
%===================    figure   =================================

In the ferromagnetic case a hard-axis anisotropy (r.h.s. of
\figref{F-7}) also has the effect to weaken the magnetization
compared to a Brillouin function. The interesting case is here
again the case of easy-axis anisotropy (l.h.s. of
\figref{F-7}). If the magnetization for a small field, here
$g\mu B/|J|=0.1$, is plotted against temperature (\figref{F-8})
one again sees that a hard-axis anisotropy tends to smear out
and weaken the magnetization (r.h.s. of \figref{F-8}) whereas
the curves appear sharper for easy-axis anisotropy (l.h.s. of
\figref{F-8}). It is worthwhile to investigate the energy
spectrum of the latter configuration in some detail and compare
it to the respective cases of intermediate and strong coupling.

%%%%%%%%%%%%%%%%%%%%%%%%%%%%%%%%%%%%%%%%%%%%%%%%%%%%%%%%%%%%%%%%%%%%%%%%
\subsection{Possible SMM behavior}

%===================    figure   =================================
\begin{figure}[ht!]
\centering
\includegraphics[clip,width=45mm]{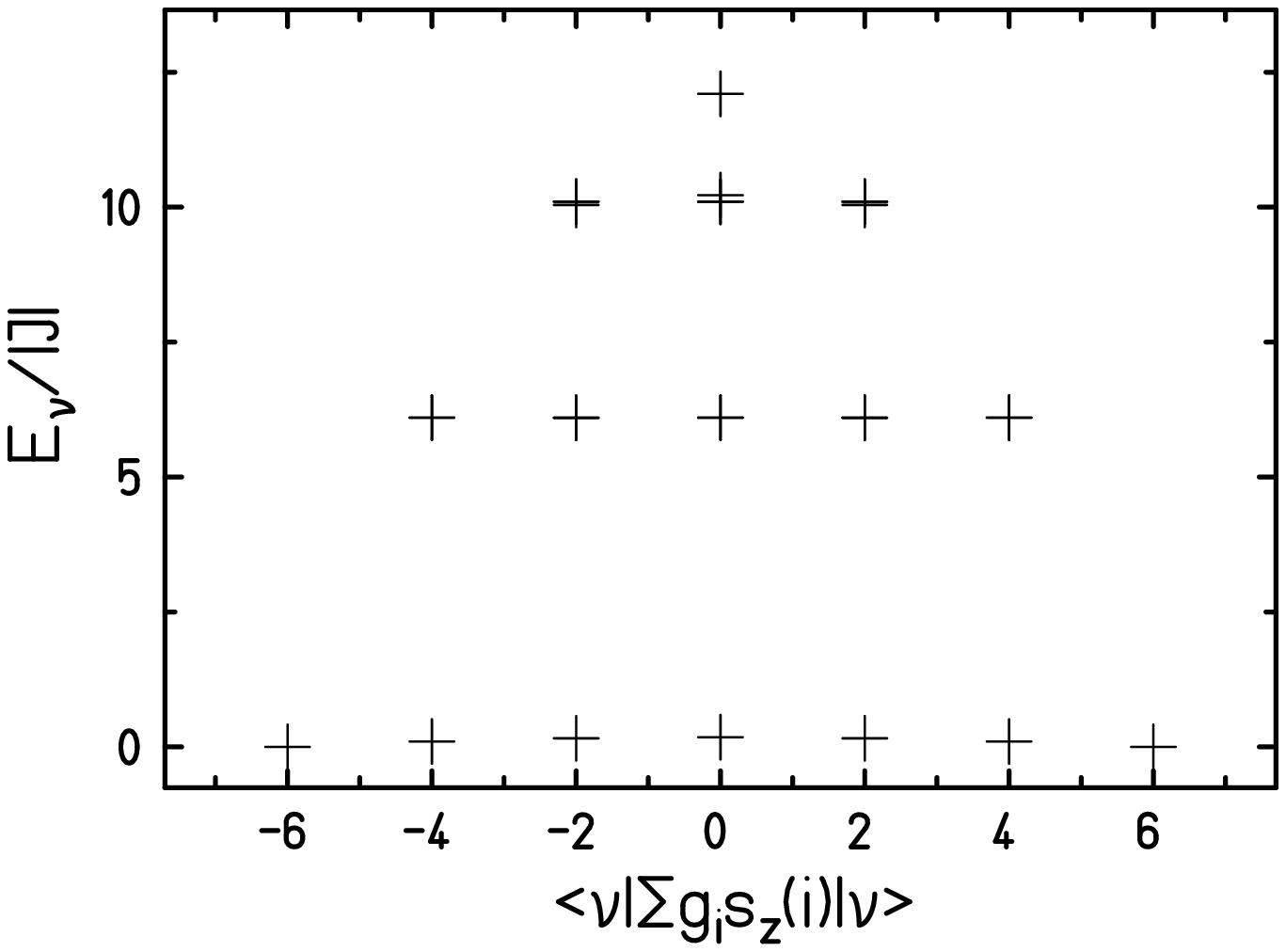}
\quad
\includegraphics[clip,width=45mm]{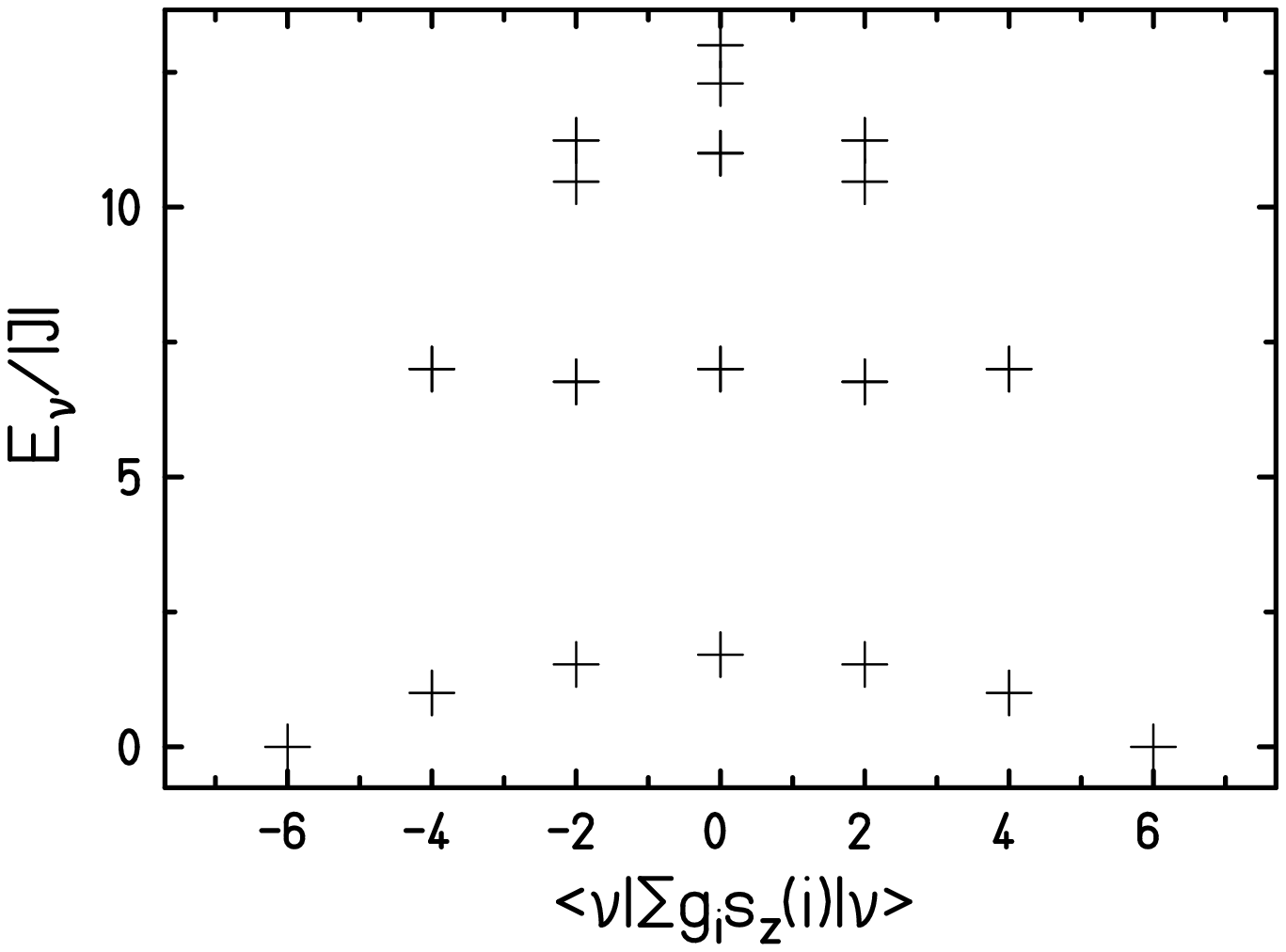}
\quad
\includegraphics[clip,width=45mm]{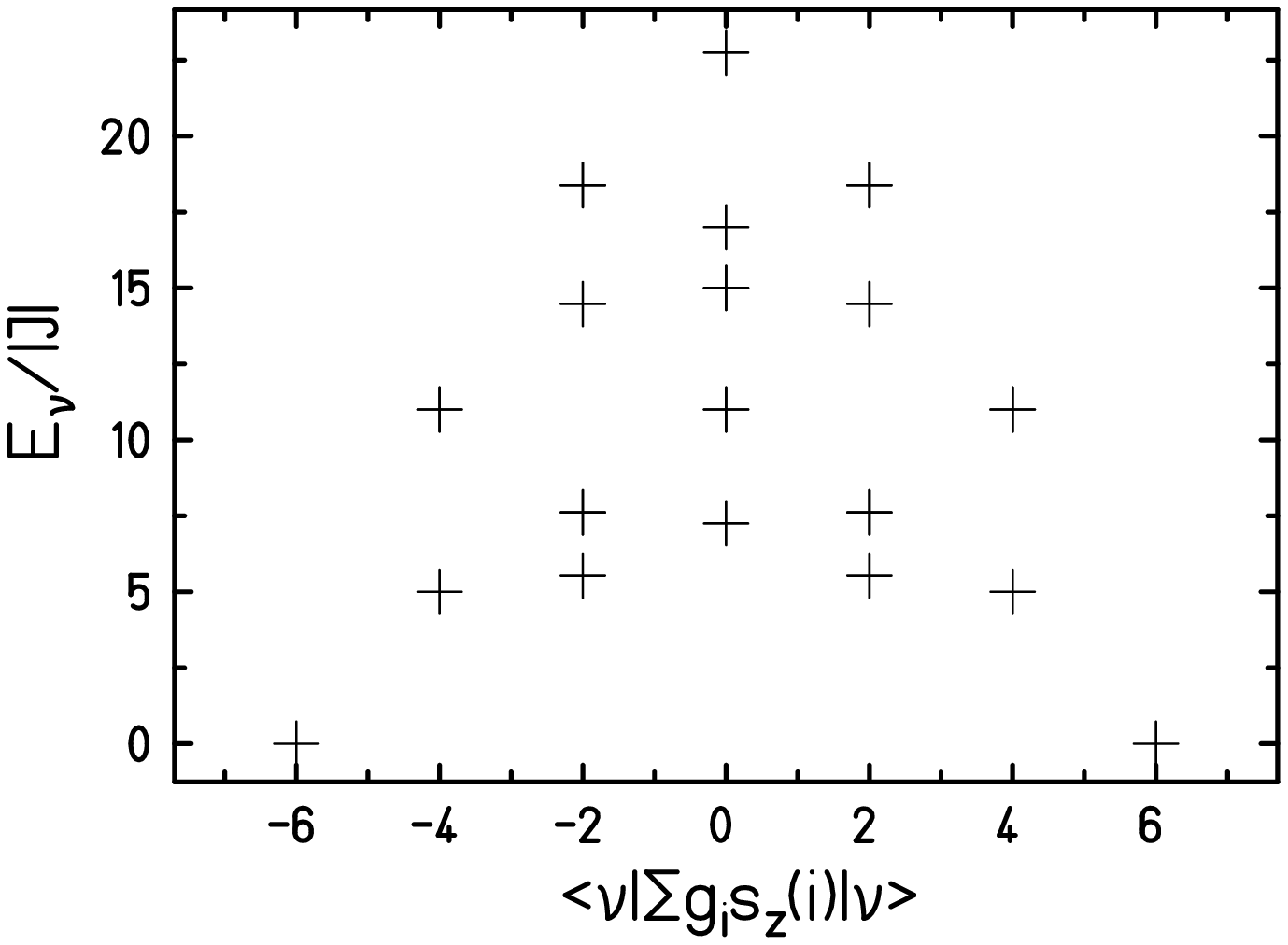}
\caption{Energy eigenvalues at $B=0$ for $J=+1$ and $d=-0.1 |J|$
  (l.h.s.), $d=-1 |J|$ (middle), and $d=-5 |J|$ (r.h.s.). $\vartheta=0$.}
\label{F-9}
\end{figure}
%===================    figure   =================================

Regarding the question which of the cases is preferential for a
good SMM the energy eigenvalues have to be studied. This is done
for the zero-field case and $J=+1$ and $d=-0.1 |J|$ (l.h.s. of
\figref{F-9}), $d=-1 |J|$ (middle of \figref{F-9}), and
$d=-5 |J|$ (r.h.s. of \figref{F-9}). The anisotropy axes are
aligned in a collinear fashion, i.e. $\vartheta=0$, in order to
act constructively. For a good SMM it would be necessary to
obtain a low-lying zero-field split multiplet with both high
total spin as well as high anisotropy barrier. In addition,
these levels should be largely separated from higher lying
levels. Looking at \figref{F-9} one notices that in the strong
coupling limit the multiplets can be well separated, but the
ground-state multiplet is practically not split. In the opposite
case of large easy-axis anisotropy the ground state multiplet is
nicely split but energetically overlaps with other levels (but
not too strong). Such a scenario was for instance discussed for the
aforementioned Mn$_6$ cluster \cite{CGS:PRL08}. It turns out
that the intermediate case, where exchange and easy-axis
anisotropy are of similar magnitude provides the best chances
for obtaining a good SMM: the ground-state multiplet is
correctly zero-field split and the higher lying states are well
separated.

%%%%%%%%%%%%%%%%%%%%%%%%%%%%%%%%%%%%%%%%%%%%%%%%%%%%%%%%%%%%%%%%%%%%%%%%
\section{Summary}

In this article the influence of non-collinear anisotropy axes
on the magnetic response of a small spin cluster was
discussed. The main result is that for good single molecule
magnets ferromagnetic coupling together with an anisotropy of
collinear easy axes that is of similar magnitude is
preferential. Various other scenarios have been discussed.

%%%%%%%%%%%%%%%%%%%%%%%%%%%%%%%%%%%%%%%%%%%%%%%%%%%%%%%%%%%%%%%%%%%%%%%%
\section*{Acknowledgement}

This work was supported by the German Science Foundation (DFG)
through the research group 945.

%%%%%%%%%%%%%%%%%%%%%%%%%%%%%%%%%%%%%%%%%%%%%%%%%%%%%%%%%%%%%%%%%%%%%%%%
%\bibliographystyle{/home/schnack/tex/bibtex/bst/fmdplain.bst}
%\bibliography{/home/schnack/tex/bibtex/js-own,/home/schnack/tex/bibtex/js-mag,/home/schnack/tex/bibtex/js-mis}

\end{document}